\documentclass[11pt,a4paper]{article}
\usepackage[T1]{fontenc}
\usepackage[utf8]{inputenc}
\usepackage[a4paper,margin=23mm,top=21mm,bottom=22mm]{geometry}
\usepackage{amsmath,amssymb,graphicx,booktabs,tabularx}
\usepackage{flafter}
\usepackage{placeins}
\usepackage[font=small,labelfont=bf]{caption}
\usepackage{xcolor}
\usepackage[colorlinks=true,allcolors=blue!45!black]{hyperref}
\usepackage[backend=bibtex,style=numeric-comp,sorting=none]{biblatex}
\graphicspath{{figures/}}
\newcommand{\BR}{\mathcal B}
\newcommand{\Emiss}{E_{\rm miss}}
\newcommand{\GeV}{\mathrm{GeV}}
\newcommand{\TeV}{\mathrm{TeV}}
\newcommand{\SM}{\mathrm{SM}}

\newcommand{\dd}{\mathrm d}
\newcommand{\gq}{\widetilde gq'}
\newcommand{\gpsi}{\widetilde gq'_\psi}
\newcommand{\lamt}{\lambda_t}
\newcommand{\eeff}{\overline\epsilon}

\begin{document}
\begin{center}
{\Large\bfseries $B\to$ invisible decays as probes of light new physics}\par\vspace{9pt}
{\large Mart\'in Novoa-Brunet}\par\vspace{4pt}
\href{mailto:mnovoa@ifae.es}{mnovoa@ifae.es}\par\vspace{4pt}
{\small Instituto de F\'isica Corpuscular, Universitat de Val\`encia-CSIC,\\
Parc Cient\'ific, E-46980 Paterna, Valencia, Spain\\
Institut de F\'isica d'Altes Energies (IFAE), The Barcelona Institute of Science and Technology, Campus UAB, 08193 Cerdanyola del Vall\`es (Barcelona)}
\end{center}
\begin{abstract}
Rare $B$ decays with missing energy probe heavy particles contributing to
$b\to s\nu\bar\nu$ and new light states produced on shell. We discuss both
possibilities in light of the Belle II excess in $B^+\to K^+\Emiss$.
Fits to the missing-mass distribution favour a two-body contribution near
$2.1\,\GeV$, while scalar and fermion pairs of masses around $0.4$-$0.6\,\GeV$ remain viable alternatives.
Extracting their couplings would profit from additional measurements, including
the $B\to K^*\Emiss$ rate and longitudinal $K^*$ polarization, together with invisible
$B_s$ decays which constrain scalar-pair interactions. We discuss how an aligned Higgsed
$U(1)'$ model can realize the single-vector explanation through a vector-like
top partner. Loop-induced electroweak mixing makes visible $Z'$ decays
unavoidable, requiring an additional invisible channel to evade dimuon
searches. We show that a light dark fermion can provide this channel and also account for
the observed dark matter abundance through resonant thermal annihilation. 
\end{abstract}

\section{Rare \texorpdfstring{$B$}{B} decays with missing energy}
\label{sec:context}
Flavour-changing neutral current (FCNC) decays provide stringent tests of
the Standard Model (SM) and its extensions. Their loop and CKM suppression
allows small new-physics (NP) contributions to compete with the SM amplitude.
The transitions $b\to s\nu\bar\nu$ are particularly interesting in this
respect: they are theoretically cleaner than $b\to s\ell^+\ell^-$ and
experimentally sensitive to any sufficiently light particles that escape
detection. Thus, the same $B\to K^{(*)}\Emiss$ measurement can probe heavy
virtual particles or the production of new light degrees of freedom.

These decays are also well suited to NP coupled predominantly to the third
generation. Existing flavour bounds strongly constrain generic interactions
with the light quarks, motivating flavour structures such as minimal flavour
violation (MFV) or approximate $U(2)^3$
symmetries~\cite{DAmbrosio:2002vsn,Barbieri:2012uh}. In these scenarios,
sizeable third-generation couplings can still induce observable effects in
$b\to s$ transitions through CKM mixing.

The Belle II collaboration has reported evidence for $B^+\to K^+\nu\bar\nu$
using $362\,\mathrm{fb}^{-1}$ of data. Combining inclusive- and hadronic-tag
analyses, and assuming an SM template for the $q^2$ distribution, they
find~\cite{Belle-II:2023esi}
\begin{equation}
 \BR(B^+\to K^+\nu\bar\nu)_{\text{SM template}}
       =(2.3\pm0.7)\times10^{-5}.
 \label{eq:measurement}
\end{equation}
This result has a significance of $3.5\sigma$ relative to the background-only
hypothesis and lies $2.7\sigma$ above the SM expectation. Interestingly, the excess is concentrated at low
reconstructed missing mass, $q^2_{\rm rec}\simeq3$-$5\,\GeV^2$. Its
distribution therefore contains important information beyond the integrated rate.
Due to efficiency and missing-mass smearing effects, which change the populations
of reconstructed bins, the extracted branching fraction can be
influenced by the assumed shape of the $q^2$ distribution.
The branching fraction in Eq.~\eqref{eq:measurement} is extracted by fitting
an assumed signal spectrum, the SM spectrum, rather than by summing independently measured
partial rates. Scalar or tensor interactions, and especially
light-particle production, can therefore substantially change
the extracted branching fraction through their different $q^2$ distributions.
While the localized excess is suggestive of light NP resonance production, 
it could also reflect an unaccounted-for background
contribution, a possibility requiring further experimental investigation.

Complementary to $B\to K\Emiss$, the $K^*$ channel probes additional quark-current
structures, while the inclusive $X_s$ rate and fully invisible $B_s$ decays
provide complementary, but less stringent, constraints. Potentially, searches with $\pi$, $\rho$ and $\phi$
final states could test the flavour structure of the
interaction~\cite{Belle:2013tnz,Belle:2017oht}. Interestingly, Belle has recently performed a
dedicated search for two-body $B\to hX_{\rm inv}$
decays~\cite{Belle:2026kum}. This search sets stringent mass-dependent limits which exclude a two-body invisible-particle interpretation of the Belle II excess in several mass ranges.
However, it vetoes the region near $m_X\simeq2\,\GeV$ favoured by the Belle II spectrum,
rather than fitting the narrow backgrounds from $B^+\to K^+\bar D^{(*)0}$ decays
whose charm decay products are not fully reconstructed.

\section{SM predictions and heavy new physics}
\label{sec:heavy}
At energies below the electroweak scale, the $b\to s\nu\bar\nu$ interaction
is described by the effective Hamiltonian
\begin{equation}
 \mathcal H_{\rm eff}=-\frac{4G_F}{\sqrt2}\lamt
       \sum_{a=L,R} C_a\mathcal O_a+\mathrm{h.c.},\qquad
 \mathcal O_{L(R)}=\frac{\alpha}{4\pi}
 (\bar s\gamma_\mu P_{L(R)}b)
 (\bar\nu\gamma^\mu(1-\gamma_5)\nu),
 \label{eq:wet}
\end{equation}
where $\lamt=V_{tb}V_{ts}^*$, $P_{L,R}=(1\mp\gamma_5)/2$, and the sum over
neutrino flavours is implicit. Only the left-handed operator contributes
in the SM, with $C_L^{\SM}=-X_t/\sin^2\theta_W\simeq-6.4$ and
$C_R^{\SM}=0$. The resulting branching fractions
are~\cite{Parrott:2022zte,Buras:2014fpa,Bolton:2025fsq}
\begin{equation}
\begin{aligned}
 \BR(B^+\to K^+\nu\bar\nu)_{\rm SD}&=(4.97\pm0.37)\times10^{-6},\\
 \BR(B^+\to K^+\nu\bar\nu)_{\rm total}&=(5.58\pm0.37)\times10^{-6},\\
 \BR(B^0\to K^{*0}\nu\bar\nu)&=(8.95\pm1.00)\times10^{-6},\\
 \BR(B\to X_s\nu\bar\nu)_{\rm SD}&=(2.9\pm0.3)\times10^{-5}.
\end{aligned}
\label{eq:smrates}
\end{equation}
Here SD denotes the short-distance contribution. The charged mode also
receives a contribution from $B^+\to\tau^+\nu_\tau$ followed by
$\tau^+\to K^+\bar\nu_\tau$, included in the second line.
Unlike $b\to s\ell^+\ell^-$, these decays are free from the nonlocal
electromagnetic charm contribution. The main uncertainties in the exclusive
predictions arise from the form factors and CKM inputs. Fully invisible
$B_s$ decays are instead extremely suppressed in the SM: the two-neutrino
rate vanishes in the massless-neutrino limit, and multi-neutrino final states
also have negligible rates at present experimental sensitivities.

Heavy NP modifies the Wilson coefficients in Eq.~\eqref{eq:wet}. For
lepton-flavour-universal vector interactions, it is convenient to
define~\cite{Buras:2014fpa}
\begin{equation}
 \epsilon^2=\frac{|C_L|^2+|C_R|^2}{|C_L^\SM|^2},\qquad
 \eta=-\frac{\mathrm{Re}(C_LC_R^*)}{|C_L|^2+|C_R|^2}.
\end{equation}
The $K$ and $K^*$ rates and the longitudinal $K^*$ polarization then satisfy
\begin{equation}
 R_K=(1-2\eta)\epsilon^2,\qquad
 R_{K^*}=(1+\kappa_\eta\eta)\epsilon^2,\qquad
 \frac{F_L}{F_L^\SM}=\frac{1+2\eta}{1+\kappa_\eta\eta},
 \label{eq:correlations}
\end{equation}
where $R_{K^{(*)}}$ is the short-distance rate normalized to its SM value
and $\kappa_\eta\simeq1.3$, depending mildly on form factors and the
integration range. The opposite interference signs allow right-handed
quark currents to enhance $B\to K$ while respecting the $K^*$ bound.
SM gauge invariance relates these interactions to $b\to s\ell^+\ell^-$.
Charged-lepton bounds restrict heavy-NP explanations, but tau-dominated
couplings can still give sizeable effects~\cite{Allwicher:2023xba,Bause:2023mfe}.

For $B\to K\nu\bar\nu$, vector interactions preserve the SM $q^2$
dependence. Scalar and tensor interactions, however, involve different form
factors and can modify the spectrum. A recent Belle II weak effective theory
(WET) analysis~\cite{Belle-II:2025lfq} includes scalar and tensor operators.
In the normalization of Eq.~\eqref{eq:wet}, these read
\begin{equation}
\begin{aligned}
 \mathcal O_{\rm SL}
 &=\frac{\alpha}{2\pi}(\overline{\nu_L^c}\nu_L)(\bar s_R b_L),\\
 \mathcal O_{\rm SR}
 &=\frac{\alpha}{2\pi}(\overline{\nu_L^c}\nu_L)(\bar s_L b_R),\\
 \mathcal O_{\rm TL}
 &=\frac{\alpha}{2\pi}(\overline{\nu_L^c}\sigma_{\mu\nu}\nu_L)
 (\bar s_R\sigma^{\mu\nu}b_L).
\end{aligned}
\label{eq:st}
\end{equation}
 Neutrino flavour indices
are suppressed and the sum over flavours is implicit. Notice that the tensor bilinear
selects an antisymmetric combination of different neutrino
flavours~\cite{Felkl:2021uxi}. These operators enter Eq.~\eqref{eq:wet}
with $a={\rm SL,SR,TL}$, and their coefficients vanish in the SM.
The normalization absorbs Belle II's common $\alpha/(2\pi)$ prefactor into
the operators. The Belle II Lagrangian convention differs in its sign convention 
from the standard one, corresponding to
$C_a=-C_a^{\mathrm{Belle\,II}}$ for the same flavour sums.
Notice that, the scalar and tensor operators are on a different footing from their vector counterparts, as they violate lepton number and arise from
SMEFT operators of dimension seven or higher when only SM fields are
retained. Scalar and tensor interactions can also arise from
dimension-six operators in the $\nu$SMEFT, with a light right-handed
neutrino $N_R$. These operators conserve lepton number for $L(N_R)=1$, but
require additional light fields, unlike the vector interactions of the
SM-field-only dimension-six SMEFT~\cite{Rosauro-Alcaraz:2024mvx}.
For active neutrinos, the distorted spectra remain broad across the physical
range. A
localized contribution such as that from a light on-shell particle cannot
be generated by the local heavy-NP operators alone.

\section{Light new particles in \texorpdfstring{$b\to s\Emiss$}{b to s + missing energy}}
\label{sec:light}
Light NP offers a different explanation of the Belle II excess. New
weakly interacting particles can be produced on shell and carry away the
missing energy, replacing the neutrino pair of the SM decay. At the $B$-meson
scale, their interactions are described by an EFT containing the new
fields explicitly. For scalar, fermion and vector fields, denoted by
$\phi$, $\psi$ and $V_\mu$, representative interactions are
$(\bar s b)\phi$, $(\bar s\gamma_\mu b)V^\mu$,
$(\bar s b)\phi^\dagger\phi$ and
$(\bar s\gamma_\mu b)(\bar\psi\gamma^\mu\psi)$, with coefficients of the
appropriate mass dimension~\cite{Bolton:2024egx,Bolton:2025fsq}.
The missing-mass distribution depends strongly on the number of emitted
particles. A two-body decay $B\to KX$ gives $q^2=m_X^2$. For a narrow
on-shell state, this contribution is monochromatic in the true missing
mass, while its reconstructed shape and width are depend on the experimental
missing-mass resolution. Production of an $X\bar X$ pair instead yields a
continuum above $q^2=4m_X^2$, whose shape also depends on the spin and
Lorentz structure of the interaction.

Missing energy can arise from stable particles, long-lived states escaping
the detector, or mediators decaying to invisible daughters. For $X$ with
lifetime $\tau_X$, boost $\beta\gamma$ and invisible branching fraction
$\BR_{\rm inv}$, the invisible probability is approximately
\begin{equation}
 \mathcal P_{X,\rm inv}=\BR_{\rm inv}
 +(1-\BR_{\rm inv})\exp[-L/(\beta\gamma c\tau_X)].
 \label{eq:invisibility}
\end{equation}
where $L$ is the detector path length and visible decays within this distance
are assumed to be detected.  The observed missing-energy rate is therefore the production
rate weighted by $\mathcal P_{X,\rm inv}$.

\begin{figure}[t]
\centering
\includegraphics[width=.45\linewidth]{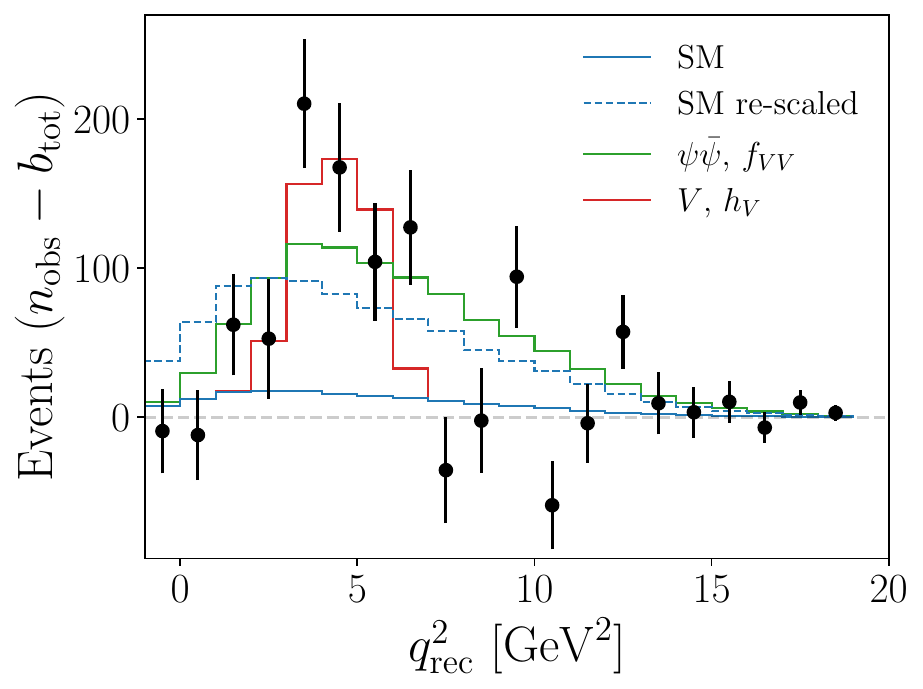}\hfill
\includegraphics[width=.49\linewidth]{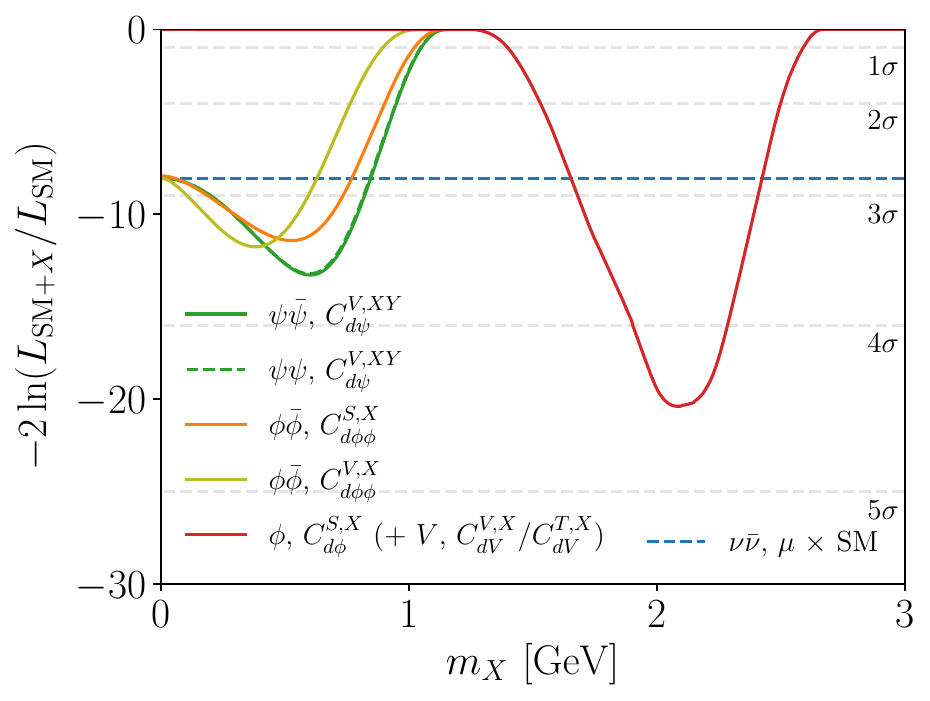}
\caption{Reconstructed missing-mass distribution (left) and profiled
mass-dependent likelihood (right) for light-particle contributions, from
Refs.~\cite{Bolton:2024egx,Bolton:2025fsq}. Single-boson emission favours
$m_X\simeq2.1\,\GeV$; pair production favours lighter invisible particles.}
\label{fig:shape}
\end{figure}

To compare the predicted spectrum with the measured distribution, we need to include
the selection efficiency and missing-mass smearing. The reconstructed
signal spectrum and binned likelihood take the form
\begin{equation}
 \begin{aligned}
 \frac{\dd N}{\dd q^2_{\mathrm{rec}}}
 &=N_{B\bar B}\int\dd q^2\,f_{q^2_{\mathrm{rec}}}(q^2)\epsilon(q^2)
       \frac{\dd\BR}{\dd q^2},\\
 \mathcal L&=\prod_i\mathrm{Pois}(n_i\mid s_i(\theta)+b_i(\theta))\,\pi(\theta).
 \end{aligned}
 \label{eq:response}
\end{equation}
where $N_{B\bar B}$ is the number of produced $B\bar B$ pairs,
$f_{q^2_{\mathrm{rec}}}(q^2)$ accounts for the smearing of reconstructed
relative to true $q^2$, and $\epsilon(q^2)$ is the selection efficiency.
Integrating $\dd N/\dd q^2_{\mathrm{rec}}$ over bin $i$ gives $s_i$.
The $n_i$ and $b_i$ denote observed and background yields, and $\pi(\theta)$
constrains the nuisance parameters.
In Refs.~\cite{Bolton:2024egx,Bolton:2025fsq}, we reconstructed binned
likelihoods using Belle II, BaBar and ALEPH results.
While several limits exist from Belle, they only provide 
integrated $K^{(*)}$ limits without a published binned $q^2$ spectra for such a
fit. We instead use BaBar's binned $B^0\to K^{*0}\Emiss$
result~\cite{BaBar:2013npw}. Its $B^0\to K^0\Emiss$ result is not included, as
its expected impact is smaller, and its efficiency information is less
detailed than that available for Belle II's charged-$K$ mode. The full
Belle II likelihood and simulated joint number densities are now
public~\cite{Belle-II:2025lfq}, allowing direct tests of new signal
distributions. 

These fits favour a single emitted boson near $2.1\,\GeV$, while
fermion pairs near $0.60\,\GeV$ and scalar pairs near $0.38$-$0.53\,\GeV$
give viable alternative shapes (Fig.~\ref{fig:shape}). Vector and spin-$3/2$
pair production gives a poorer fit for the interactions considered in
Ref.~\cite{Bolton:2024egx}.
A subsequent analysis using the public Belle II likelihood finds
$m_X=2.1^{+0.2}_{-0.1}\,\GeV$ and
$\BR(B^+\to K^+X)\mathcal P_{X,\rm inv}
=9.2^{+1.8}_{-3.4}\times10^{-6}$ for its narrow-width benchmark, with a
reported $3.0\sigma$ preference over the SM in its frequentist
test~\cite{Gartner:2026bkx}.

\section{Complementarity of invisible \texorpdfstring{$B$}{B} decays}
\label{sec:complementarity}
\begin{figure}[t]
\centering\includegraphics[width=.92\linewidth]{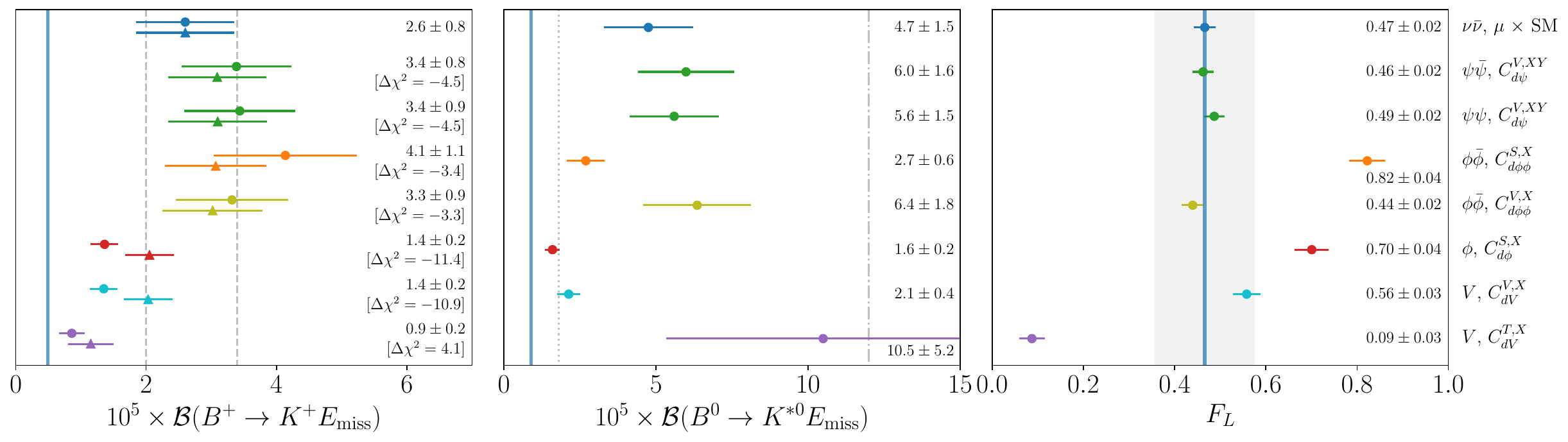}
\caption{Best-fit branching fractions for $B\to K^{(*)}\Emiss$ and
longitudinal $K^*$ polarization in the light-particle scenarios of
Ref.~\cite{Bolton:2025fsq}, compared with a rescaled SM contribution.
Triangles in the $K$ panel include efficiency and smearing and are rescaled
according to Eq.~\eqref{eq:apparent}. Their difference from the rescaled-SM
band illustrates the dependence of the extracted rate on the $q^2$ shape.}
\label{fig:rates}
\end{figure}

Using the SM-based branching fraction to constrain light NP can give an
incorrect mass or coupling, even after correcting the total efficiency.
For a normalized spectrum $p_X(q^2)$, one can define an effective signal efficiency
$\eeff_X=\int_{\rm fit}\dd q^2_{\mathrm{rec}}\int\dd q^2\,
f_{q^2_{\mathrm{rec}}}(q^2)\epsilon(q^2)p_X(q^2)$,
where the outer integral covers the fitted region in reconstructed $q^2$.
Equating the total expected signal yield under hypothesis $X$ to that of an
SM-shaped template, without using the distribution among bins, gives
\begin{equation}
 \BR_{\rm apparent}^{X}
     =\BR_{\rm physical}^{X}\frac{\eeff_X}{\eeff_\SM}.
 \label{eq:apparent}
\end{equation}
The triangles in Fig.~\ref{fig:rates} include this rescaling, with efficiency
and smearing, but remain displaced from the rescaled-SM band. The fit also
depends on the bin populations and their backgrounds: correcting the effective
signal efficiency does not recover the mass and coupling inferred from the full
$q^2$ distribution.

The pseudoscalar and axial-vector quark currents have vanishing $B\to K$
matrix elements, but contribute to $B\to K^*\Emiss$. Their coefficients
are independent in a parity basis, whereas a single chiral operator fixes
the relative parity-even and parity-odd couplings. The $K^*$ bound can
therefore exclude a chiral interaction allowed when both parity components
vary independently.

Fully invisible $B_s$ decays provide particularly useful constraints on
scalar-pair interactions~\cite{Bolton:2024egx}. For the vector/axial fermion
interactions considered here, the $B_s\to\psi\bar\psi$ rate is helicity
suppressed~\cite{Bolton:2025fsq}.

The $K^*\to K\pi$ decay also measures the longitudinal fraction through
\begin{equation}
 \frac{1}{\dd\Gamma/\dd q^2}
 \frac{\dd^2\Gamma}{\dd q^2\dd\cos\theta_K}
 =\frac34(1-F_L)\sin^2\theta_K+\frac32F_L\cos^2\theta_K,
 \qquad
 F_L(q^2)=\frac{\dd\Gamma_L/\dd q^2}{\dd\Gamma/\dd q^2}.
 \label{eq:fl}
\end{equation}
Here $\theta_K$ is the angle of the kaon in the $K^*$ rest frame relative
to the $K^*$ flight direction in the $B$ rest frame. Scenarios giving similar $K$ spectra can predict very
different $K^*$ rates and polarization~\cite{Bolton:2025fsq}.
For a $q^2$ bin $\Delta$, the polarization is the ratio of integrated
rates, $F_L^\Delta=\int_\Delta\dd\Gamma_L/\int_\Delta\dd\Gamma$.
To compare with the reconstructed angular distribution, the longitudinal
and transverse rates must be folded separately with the angular acceptance
and missing-mass resolution. This is particularly important for a two-body decay, whose
contribution can be diluted over a wide reconstructed bin.

\section{An aligned \texorpdfstring{$U(1)'$}{U(1)'} model}
\label{sec:uv}
Several UV completions of the interactions discussed above have been considered in the
literature, including axion-like particles, scalar portals, non-universal
$Z'$ models and dark photons~\cite{Altmannshofer:2023hkn,Hati:2024ppg,Gabrielli:2024wys}.
We discuss one of these, an aligned Higgsed $U(1)'$ model proposed in
Ref.~\cite{Bolton:2025lnb}, which realizes the single-vector explanation
of the Belle II excess. The model introduces a vector-like quark
$T'\sim(\mathbf3,\mathbf1,2/3,q')$ and a singlet scalar
$\Phi\sim(\mathbf1,\mathbf1,0,q')$, with
\begin{equation}
 \mathcal L\supset\bar T'(i\!\not\!D-M_T)T'
 -[y_T^r\bar T'_L\Phi u'_{Rr}+\mathrm{h.c.}]
 +|D_\mu\Phi|^2-V(H,\Phi)
 -\frac{\epsilon_B}{2}B_{\mu\nu}B'^{\mu\nu},
 \label{eq:model}
\end{equation}
where $\langle\Phi\rangle=\widetilde v/\sqrt2$ breaks the new gauge symmetry
and gives $M_{Z'}=\gq\,\widetilde v$. We take $y_T^{u,c}=0$ and
$y_T^t\ne0$, so that $T'$ mixes only with the top quark. The resulting
$W$-$t/T$ penguins generate $b\to sZ'$ with the CKM factor
$V_{ts}^*V_{tb}$. This top alignment naturally gives an MFV-like flavour
structure, with the largest FCNC effects in $b\to s$ transitions.

The top partner also induces kinetic mixing between the new vector and
hypercharge. Taking this mixing to vanish at a high scale $\Lambda$, the
leading logarithmic contribution to photon-$Z'$ mixing is
\begin{equation}
 \epsilon_A\simeq\frac{e\gq}{6\pi^2}L_T,
 \qquad L_T=\ln\frac{\Lambda^2}{m_T^2}.
 \label{eq:mixing}
\end{equation}
For $\Lambda=10^{16}\,\GeV$ and $m_T=10\,\TeV$, $L_T\simeq55$.
Mixing in the top sector further induces $Z$-$Z'$ mass mixing, which
cannot be removed by tuning the hypercharge kinetic-mixing counterterm.
Consequently, the $Z'$ couples to charged leptons and decays visibly.
Although the minimal model can reproduce the required $B\to KZ'$ production
rate, it is excluded by LHCb searches for prompt and displaced dimuon
resonances~\cite{LHCb:2015nkv,LHCb:2016awg}.

A minimal extension introduces a light vector-like SM singlet $\psi$ with
$U(1)'$ charge $q'_\psi$, opening the decay
\begin{equation}
 \Gamma(Z'\to\psi\bar\psi)=\frac{M_{Z'}}{12\pi}(\gpsi)^2
 (1+2z)\sqrt{1-4z},\qquad z=\frac{m_\psi^2}{M_{Z'}^2},
 \quad r_\psi=\frac{q'_\psi}{q'}.
 \label{eq:width}
\end{equation}
For $2m_\psi<M_{Z'}$, this channel can dominate the total width and suppress
the dimuon branching fraction sufficiently to evade LHCb. For
$M_{Z'}=2.1\,\GeV$ and $m_\psi\simeq1\,\GeV$, this is achieved with the
representative charge ratio $r_\psi=12$. Several light dark fermions with
order-one charge ratios could replace this single large charge.

\begin{figure}[t]
\centering
\includegraphics[width=.49\linewidth]{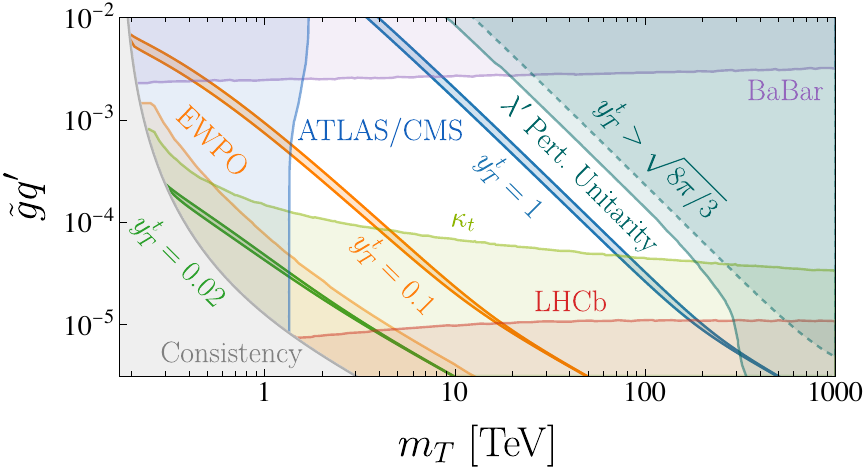}\hfill
\includegraphics[width=.47\linewidth]{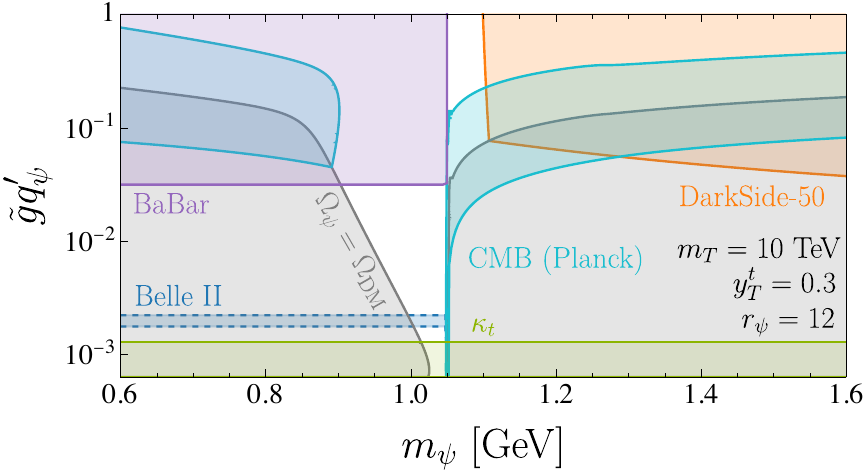}
\caption{Parameter space of the aligned $U(1)'$ model with a light dark
fermion, from Ref.~\cite{Bolton:2025lnb}. Left: the $B\to K\Emiss$ region
and constraints from flavour, collider and electroweak observables, and
perturbative unitarity, for $r_\psi=12$ and $m_\psi=1\,\GeV$.
Right: dark-matter constraints for
$m_T=10\,\TeV$, $y_T^t=0.3$ and $r_\psi=12$. The gray curve gives the
observed relic abundance calculated with \texttt{micrOMEGAs}~\cite{Alguero:2023zol}; the cyan region is
excluded by energy injection into the CMB from late-time annihilation.}
\label{fig:uv}
\end{figure}

The surviving parameter space in Fig.~\ref{fig:uv} is further constrained
by BaBar monophoton searches, direct top-partner production, the modified
top Yukawa coupling and electroweak precision
observables~\cite{BaBar:2017tiz,CMS:2020mpn}. The scalar-sector
perturbative-unitarity bound assumes a vanishing Higgs-portal quartic
$\lambda'(\Lambda)=0$ at $\Lambda=10^{16}\,\GeV$, with its low-energy
value generated radiatively, and a scalar mass $M_\phi>M_h$.

\subsection{Resonant dark matter}
The dark fermion can also account for the observed DM abundance if it is
stable on cosmological scales. Assuming an initial thermal population,
its abundance is depleted by $\psi\bar\psi\to Z'^*\to\SM$. For
$m_\psi<M_{Z'}/2$, a sufficient annihilation rate can be obtained near
$2m_\psi\simeq M_{Z'}$. The $s$-wave-dominated thermal average near the
narrow resonance is, at leading order in $\epsilon_\psi>0$,
\begin{equation}
 \epsilon_\psi=\frac{M_{Z'}^2-4m_\psi^2}{4m_\psi^2},\qquad
 \langle\sigma v\rangle\simeq
 \frac{12\pi^{3/2}}{m_\psi^2}\gamma_{Z'}
 \BR_{Z'}\,x^{3/2}e^{-x\epsilon_\psi},
 \label{eq:resonance}
\end{equation}
where $x=m_\psi/T$, $\gamma_{Z'}=M_{Z'}\Gamma_{Z'}/(4m_\psi^2)$ and
$\BR_{Z'}=\BR_{\psi\bar\psi}\BR_\SM$. Here
$\BR_{\psi\bar\psi}=\Gamma(Z'\to\psi\bar\psi)/\Gamma_{Z'}$ and
$\BR_\SM=1-\BR_{\psi\bar\psi}$ includes all SM final states.
When $\BR_{\psi\bar\psi}\simeq1$ and kinetic mixing dominates the
coupling to SM states, the approximate relic abundance is
independent of $r_\psi$. The numerical calculation finds a viable region at
$0.9\lesssim m_\psi/\GeV\lesssim1.0$.
Closer to the pole, annihilation can continue after kinetic decoupling;
in this regime a calculation assuming a common dark-sector and SM
temperature is no longer reliable.

The Planck determination of the DM density requires
$\Omega_\psi h^2\leq\Omega_{\rm DM}h^2\simeq0.12$, with equality if $\psi$
constitutes all the DM. The late-time-annihilation constraint from Planck
is parametrized by
\begin{equation}
 p_{\rm ann}=\sum_i\frac{f_i^{\rm eff}\langle\sigma v\rangle_i}{m_\psi}
        \left(\frac{\Omega_\psi}{\Omega_{\rm DM}}\right)^2,
 \label{eq:cmb}
\end{equation}
where $f_i^{\rm eff}$ is the energy-deposition efficiency for channel
$i$~\cite{Planck:2018vyg}. The direct-detection limit is instead rescaled
by $\Omega_\psi/\Omega_{\rm DM}$ for a subdominant relic. The DarkSide-50 constraint in
Fig.~\ref{fig:uv} arises from scattering on argon through the predominantly
electromagnetic coupling induced by mixing~\cite{DarkSide:2018bpj}.
Electron-scattering bounds also leave the resonant region allowed.

Extensions with additional scalar fields and different $U(1)'$ charges
for the left- and right-handed dark fermions can yield Majorana states with axial-vector $Z'$
couplings and $p$-wave-suppressed annihilation into SM states, potentially
relaxing the CMB and direct-detection bounds. Heavier DM can instead
annihilate to $Z'Z'$. In that case, lighter dark states are
needed to provide the invisible $Z'$ decays responsible for Belle II.

\section{Conclusions}
The Belle II excess can arise from heavy NP in $b\to s\nu\bar\nu$ or
from new light particles. The missing-mass distribution favours a two-body
decay with a boson near $2.1\,\GeV$, while scalar and fermion pairs remain
viable. Their masses and couplings require a differential analysis as
different $q^2$ shapes affect both the efficiency and the inferred rate.
The public Belle II likelihood now allows these possibilities to be
tested directly.

The $K^*$ rate and longitudinal polarization distinguish interactions that
are degenerate in $B\to K\Emiss$, while inclusive and fully invisible
decays provide complementary constraints. In the aligned $U(1)'$ model,
a vector-like top partner generates the required FCNC coupling, but also
induces visible $Z'$ decays excluded by LHCb. Adding a light dark fermion
allows a dominant invisible width and the observed DM abundance through
resonant freeze-out. Improved missing-mass spectra and $K^*$ polarization
will be particularly important in testing these explanations.

\section*{Acknowledgements}
I thank Patrick D. Bolton, Svjetlana Fajfer and Jernej F. Kamenik for the
collaborations on which this contribution is based, and the organizers of
Beauty 2026 for the invitation. I also acknowledge financial support from the Spanish Ministry of Science,
 Innovation and Universities (Agencia Estatal de Investigaci\'on MICIU/AEI/10.13039/501100011033) through the Severo Ochoa Centers of Excellence Programme under Grants No. CEX2024-001441-S and No.~CEX2023-001292-S, and through Grants No.~PID2023-146142NB-I00
and No. PID2023-146220NB-I00.

\printbibliography
\end{document}